\documentclass[journal]{vgtc}                     

\onlineid{0}

\vgtccategory{Research}

\title{Beyond Score-Based Gamification: Designing Spatiotemporal and Musical Experiences for VR Neck Rehabilitation}

\author{%
    \authororcid{David Perron}{0009-0005-3713-447X},
    \authororcid{Pascal Spiegler}{0009-0002-6600-957X},
    Haitham Abdel-Salam,
    \authororcid{Chanelle Montpetit}{0000-0001-8389-5003},\\
    \authororcid{Gabriel Vigliensoni}{0000-0003-0274-4356},
    \authororcid{Maryse Fortin}{0000-0002-1189-3591}, and
    \authororcid{Yiming Xiao}{0000-0002-0962-3525}
}

\authorfooter{
\item
David Perron, Pascal Spiegler, and Yiming Xiao are with the
Department of Computer Science and Software Engineering,
Concordia University, Montreal, QC H3G 1M8, Canada.
E-mail: pe\_dav@live.concordia.ca, pascal.spiegler@mail.concordia.ca, 
yiming.xiao@concordia.ca.

\item
Haitham Abdel-Salam is with the
Department of Electrical and Computer Engineering,
Concordia University, Montreal, QC H3G 1M8, Canada.
E-mail: haithamhany7@gmail.com.

\item
Chanelle Montpetit and Maryse Fortin are with the
Department of Health, Kinesiology and Applied Physiology,
Concordia University, Montreal, QC H3G 1M8, Canada.
E-mail: {c\_montp, maryse.fortin}@concordia.ca.

\item
Gabriel Vigliensoni is with the
Department of Design and Computation Arts,
Concordia University, Montreal, QC H3G 1M8, Canada.
E-mail: gabriel.vigliensoni@concordia.ca.
}

\abstract{%
  Pain-related anxiety and fear of movement are major barriers to adherence and therapeutic outcomes in rehabilitation exercises for chronic neck pain. Virtual reality (VR) enables the design of immersive experiences that can transform repetitive therapeutic movements into engaging and emotionally supportive interactions. In this \textcolor{black}{exploratory} work, we investigate how experience-oriented gamification can reduce anxiety and improve user experience during VR-based neck range-of-motion (ROM) exercises. We introduce two novel interaction paradigms that embed therapeutic neck movements within multisensory VR experiences. The first paradigm, Spatiotemporal Progression, couples head-tracked trajectories with environmental progression in a tropical island setting, where movement segments dynamically transform time of day, weather, and spatial location as experiential rewards. The second paradigm, Musical Interaction, maps movement segments to meditative musical notes layered with relaxing ambient soundscapes. We evaluate these designs against a conventional score-based gamification baseline in a controlled user study with 20 \textcolor{black}{non-patient} participants. We assess usability and user experience through subjective measures, exercise performance with motion tracking, and anxiety modulation using the Subjective Units of Distress Scale (SUDS), heart rate, and skin conductance. Our findings \textcolor{black}{in the non-patient cohort} suggest that, in comparison with traditional score-based gamification design, immersive environmental and musical feedback show better potential to reduce anxiety and improve user experience, with little to no impact on successful performance of the exercise. Our \textcolor{black}{preliminary} results highlight the \textcolor{black}{potential} value of experience-based interaction design for VR rehabilitation, suggesting an alternative to performance-centric gamification that prioritizes emotional engagement without compromising therapeutic efficacy.
}


\teaser{
  \centering
  \includegraphics[width=\linewidth,
        alt={Overview of the three VR rehabilitation paradigms and the neck range-of-motion exercise.}]{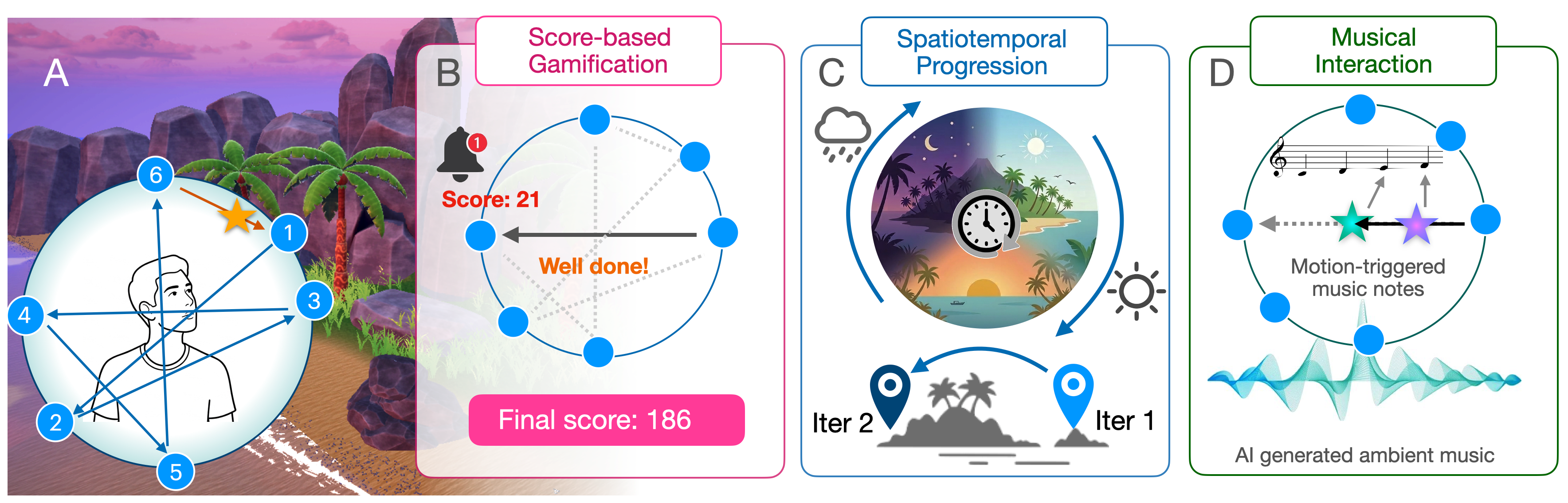}
  \caption{Demonstration of different game designs for the neck range-of-motion (ROM) exercise. A. Illustration of the ROM exercise routine with steady motion between six personalized calibration points, set within a tropical island virtual environment with associated soundscapes (e.g., tides and birds). B. Mechanism of the score-based game design, where the score for each movement segment is displayed alongside encouragement text. C. Spatiotemporal progression game design, where the time of day and weather in the virtual environment change when the user holds a calibration point, and a location teleportation is triggered at the start of each new exercise iteration. D. Musical interaction game design, where musical notes are triggered with neck movements in a segment on top of an AI-generated ambient music.}
  \label{fig:teaser}
}

\graphicspath{{figs/}{figures/}{pictures/}{images/}{./}}

\usepackage{booktabs}                  
\usepackage{lipsum}                    
\usepackage{mwe}                       
\usepackage{ccicons}                   

\usepackage{mathptmx}                  

\begin{document}

\firstsection{Introduction}

\maketitle

Chronic neck pain is one of the most prevalent musculoskeletal disorders worldwide and a major contributor to long-term disability, affecting millions of individuals \cite{Kazeminasab:2022:NPG}. It is often associated with a lack of physical activity, poor posture, and musculoskeletal imbalances, which are increasingly common in the digital age \cite{Steilen2014}.
\textcolor{black}{Although chronic neck pain commonly affects middle-aged adults, especially in the female population, it is also increasingly impacting the young adults \cite{wang2026global,murray2022prevalence,gao2023risk}.} Beyond physical impairments, such as reduced cervical mobility and muscle stiffness, chronic neck pain is also strongly associated with psychological factors, particularly pain-related anxiety. This anxiety arises from anticipation of pain triggered by physical activity or fear of re-injury, and over time leads to fear of movement (kinesiophobia) \cite{elbinoune2016chronic}, which further reduces physical activity and contributes to long-term disability \cite{vlaeyen2016fear}. Exercise therapies have been shown to improve both pain levels and musculoskeletal function \cite{Barreto2019}. However, in addition to the tedious nature and accessibility challenges of such therapy \cite{Himler2023}, associated anxiety can significantly reduce adherence to prescribed rehabilitation exercises and limit therapeutic outcomes \cite{asiri2021kinesiophobia}. Therefore, rehabilitation approaches for chronic neck pain must address both the biomechanical aspects of cervical movement and the psychological factors that influence how patients perceive and perform therapeutic exercises.

With increasing affordability, virtual reality (VR) has emerged as a promising technology for physical rehabilitation (e.g., stroke recovery), where therapeutic exercises are embedded within immersive and interactive games \cite{weber2020commercially} to enhance patient engagement, adherence to the program, and accessibility to therapy. Unlike typical exercise games (exergames) that focus on maximizing exertion, gamified physical rehabilitation emphasizes controlled, precise, and sometimes nuanced movements \cite{Chiu2024} instead. While most existing VR rehabilitation applications target the limbs, those tailored for \textit{neck pain rehabilitation} remain largely under-explored, despite a growing patient population and current constraints on clinical resources \cite{Global2024}. Among the limited work in this area, Orr et al. \cite{Orr2023} conducted a retrospective study on VR-based rehabilitation for low back and neck pain, where patients used a commercial VR game \textit{Rotate} from the company \textit{XRHealth} to perform neck range-of-motion exercise, confirming the feasibility and safety of VR-assisted neck exercise therapy. Similarly, Guo et al. \cite{Guo2024} performed a randomized controlled trial with chronic neck pain patients using commercial VR rehabilitation games incorporating shooting and token collection mechanics to guide cervical movements. Their findings showed improvements in patient satisfaction, symptom relief, and engagement with rehabilitation exercises. Additionally, recent systematic reviews have also shown that VR-based interventions can improve pain, cervical mobility, and functional outcomes for neck pain patients \cite{hao2024virtual,ye2023use}. Despite the demonstrated clinical benefits, existing VR cervical training systems commonly employ score-based gamification with the main goals of exercise motivation and performance feedback, often overlooking psychological factors (pain-related anxiety and kinesiophobia) that influence patients’ willingness to perform therapeutic movements. Notably, prior research has shown that performance score-based gamification mechanisms (e.g., point/token collection) can introduce evaluative stress \cite{yang2021understanding} and are rarely used for digital stress management applications \cite{hoffmann2017gamification}.

Beyond physical rehabilitation, VR-based relaxation interventions have been shown to significantly lower anxiety using validated subjective measures, such as the State-Trait Anxiety Inventory (STAI) and the Subjective Units of Distress Scale (SUDS), as well as physiological markers (e.g., heart rate, skin conductance, and EEG) \cite{riches2023virtual,xu2024effectiveness}. Systematic reviews \cite{savoric2025systematic} have shown that reductions in perceived stress and high engagement in immersive VR occur across both clinical and general populations, and the rendering of natural environments has been commonly adopted. Additionally, recent experimental studies \cite{zhang2026vr,zhang2026asafeplace, skiers2025portable} further emphasize personalized and multi-sensory experiences, including user-controlled virtual scenes, interactive breathing exercises, and tangible interfaces, which reduce both self-reported and physiological stress. Furthermore, research also demonstrates that the degree of immersion in VR simulations could modulate anxiety reduction \cite{hosseini2025evaluating} when it comes to exposure therapy-based intervention. \textit{Beyond visual immersion}, musical and auditory interaction in VR has been shown to support relaxation. For example, Hsieh at al. \cite{hsieh2023effect} show that appropriate selection of accompanied soundscapes can enhance restorative natural environment simulation in the VR (e.g., water sound in a forest environment), calming emotions and relieving anxiety. On the other hand, active VR music games, music-synchronized environments, and interactive sound interfaces could reduce anxiety and modulate physiological arousal \cite{kim2025designing, lecamwasam2023investigating,yang2025analgesic}. Together, these previous studies offer strong empirical support for experience-centered VR design to lower anxiety and stress in therapeutic and rehabilitative applications.

Despite advances in VR-based rehabilitation and anxiety reduction, several gaps still remain for VR-facilitated chronic neck pain interventions. \textbf{First}, at the clinical application level, most systems focus only on biomechanical performance (e.g., accuracy and task completion), largely overlooking anxiety reduction and fear-of-movement management. \textbf{Second}, for the sensory designs, the potential of multi-sensory immersive feedback, such as dynamic environmental changes or music-synchronized cues, to support both relaxation and motor engagement remains unexplored. \textbf{Finally}, in terms of reward mechanisms, current approaches rely on generic score-based gamification, lacking effective designs that combine emotional comfort, motivation, and biomechanical guidance. These gaps raise a key question: \textit{How do VR interaction paradigms influence user experience, motivation, and emotional response during therapeutic neck exercises?} Even though gamification is widely used to make repetitive tasks engaging, little is known about how different gameplay structures affect perception and anxiety in cervical rehabilitation. Understanding this is instrumental for immersive systems research, where interaction design could directly shape behavior and emotional state.

This \textcolor{black}{exploratory study aims to} address gaps in VR-based neck pain rehabilitation by integrating anxiety reduction with therapeutic movements in a staple range-of-motion (ROM) exercise, designed by a therapist. Based on established theoretical frameworks in psychology and neuroscience \cite{koelsch2014brain,grahn2007rhythm,ulrich1991stress,barsalou2008grounded,juslin2011handbook}, we propose two experience-centered game paradigms: \textbf{spatiotemporal progression}, which links neck movements to dynamic natural environmental changes, and \textbf{musical interaction}, which synchronizes motion with meditative auditory cues. Both designs leverage multi-sensory immersive feedback as implicit rewards to enhance engagement and emotional regulation, in contrast to conventional \textit{score-based gamification}. Additionally, we explored the use of generative AI for soundscape composition within the musical interaction paradigm. Using subjective anxiety ratings (SUDS), physiological measures (heart rate and skin conductance), exercise performance metrics, and user experience (UX) questionnaires in a user study of 20 \textcolor{black}{non-patient} participants, we evaluate how interaction paradigms affect stress, performance, and motivation, providing empirical evidence to guide experience-centered VR rehabilitation design.

\section{Hypotheses}
\label{sec:hypotheses}
To investigate the impact of our experience-based game designs for neck rehabilitation exercises, we conducted a user study comparing the two proposed experience-based game designs (spatiotemporal progression and musical interaction) against conventional score-based gamification. We formulate the following hypotheses:

\noindent
\textbf{H1:} Experience-based game designs reduce anxiety and stress more effectively than conventional score-based gamification during the ROM neck rehabilitation exercise;

\noindent
\textbf{H2:} Despite the use of implicit rewards, experience-based designs yield exercise performance quality (measured by movement accuracy) comparable to score-based gamification explicit rewards;

\noindent
\textbf{H3:} Compared with score-based gamification, experience-based game designs offer a more positive perceived user experience for the embedded cervical rehabilitation exercise.

\section{Methods and Materials}

\subsection{Hardware and Software Setup}
We implemented our VR application using Unreal Game Engine 5.6.1, along with Niagara and MetaSound to produce the visuals and sounds. In addition, we used Google's MusicFX AI tool \textcolor{black}{and Gemini 2.5} to generate ambient music \textcolor{black}{the composition of the musical notes} for the Musical Interaction paradigm. A Meta Quest 3 VR headset was used for the development and user study, during which it was connected to a desktop PC (RTX 2080 Super GPU, Intel i9 CPU 3.70 GHz, 32 GB RAM) via Quest Link. In our user study, we collected participants' physiological data using the EmotiBit wearable biometric sensor module \cite{montgomery2023introducing}, which was connected to a Lenovo Legion 7i laptop (RTX 5070Ti, 32 GB of RAM, Intel CPU) for data streaming via a local network. Participants wore the EmotiBit on their left index finger using an adjustable strap while using the right controller to interact with the VR application; the left controller was not required during gameplay.

\subsection{Virtual Environment Design and Implementation}

Following Bratman et al. \cite{bratman2015nature}, who establish links between mental well-being and nature experiences, we designed a virtual environment of a small tropical island surrounded by an ocean, which offers a diverse range of audiovisual elements. While the environment adopts relatively low-polygon 3D models for the scene objects (e.g., rocks, palm trees, and foliage) for rendering efficiency, we employed a realistic dynamic sky asset to deliver an immersive experience of time-of-day and weather changes (e.g., rain, sunshine, and cloudy conditions), along with dynamically shifting sunlight direction and evening stars during runtime. In addition, the soundscape incorporates ocean waves, birds, and insects \cite{hsieh2023effect}. The user can traverse different locations of the island while experiencing changing times of day and weather conditions. Finally, post-processing effects, including bloom and atmospheric fog, were applied to enhance the immersive, dream-like experience.

\subsection{Exercise Design and General Game Play}
\begin{figure}
    \centering
    \includegraphics[width=1\linewidth]{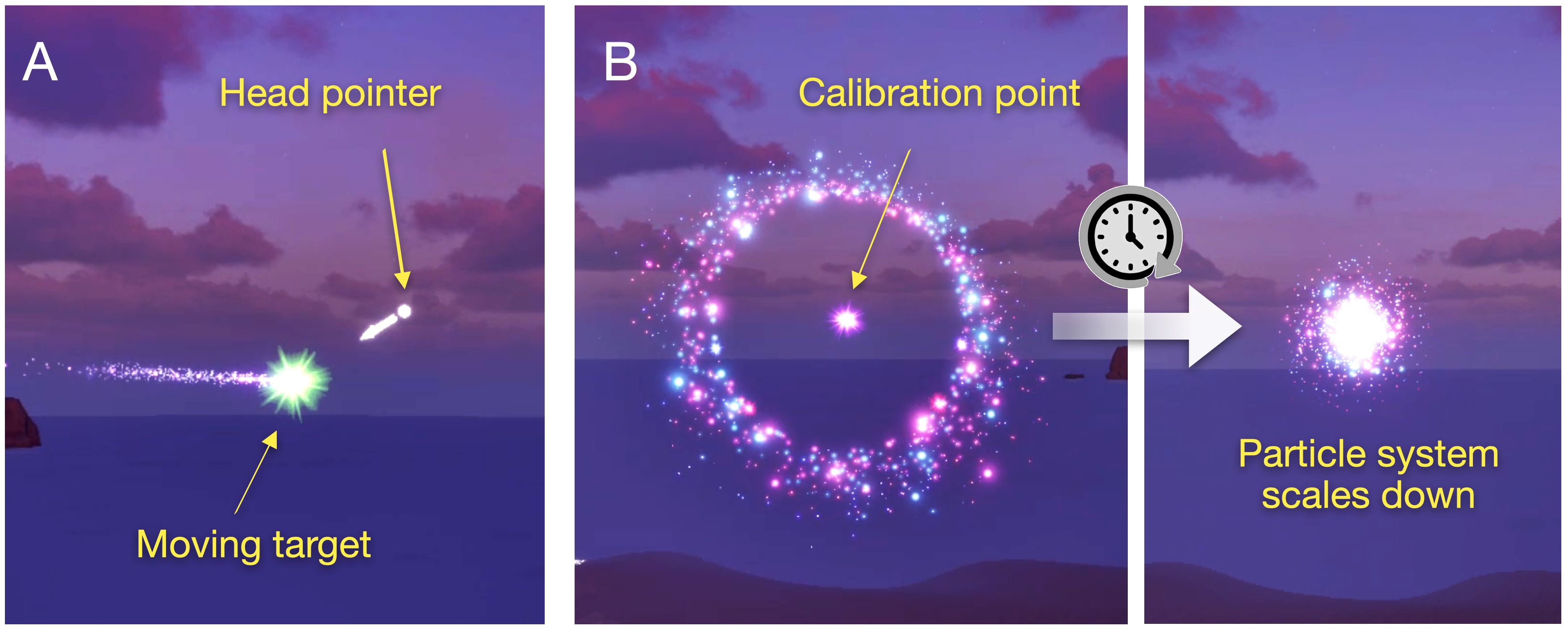}
    \caption{Common visual elements across all exercise games. A. Visual cues during movement: the moving target guides the neck movement, while the head pointer shows the current head-pointing direction; the white arrow indicates the level of deviation from the target. B. During fixation at a calibration point, a light particle animation is used as time progression feedback.}
    \label{fig:lightsystem}
\end{figure}

For neck pain rehabilitation, range-of-motion (ROM) exercises help restore cervical mobility through active movement of the neck in its natural directions of motion, such as extension (head lift), flexion (chin drop towards the chest) and rotation (turning the head to the side). The exercise incorporated in the study was designed by a certified athletic therapist collaborating on the research project. Specifically, in the dynamic tropical island environment, the user follows a luminous starlight with head pointing across various individually calibrated points for the neck motion range to enable controlled, smooth motions of flexion, extension, and rotation, as shown in Fig. \ref{fig:lightsystem}A. The exercise protocol requires the user to complete two iterations of the full movement set (flexion, extension, and rotation), with each round following the order of the calibration points 1$\sim$6 (Fig. 1A). At each calibration point, the user will fix their pose for 9 seconds to stretch their neck muscles. 

For the general game play in the designed virtual environment, we designed two dynamic visual features to facilitate the user during the exercise. \textit{First}, during the smooth neck movement between calibration points, as shown in Fig. \ref{fig:lightsystem}A, a semi-transparent white dot (head pointer) indicates the current head-pointing direction, which the user should align with the moving starlight guide. When the head pointing deviates from the guide, the white dot will change into an arrow that points to towards the target, with the arrow stretching longer as the deviation becomes more prominent. This intends to offer real-time feedback for the precision of the exercise movement. \textit{Second}, to indicate time progression and mitigate the sense of boredom during neck pose fixation at calibration points, two colorful rings of sparkling lights centered at the current calibration point will appear and gradually shrink down until they converge, accompanied by a light bell strike when the fixation period ends (Fig. \ref{fig:lightsystem}B). At the calibration points, the user is instructed to align the head pointer with the starlight guide. If a misalignment beyond a threshold occurs, the fixation time will be prolonged and the light ring animation pauses.

\subsection{Personalized Calibration and Performance Score}

\subsubsection{Personalized Movement Calibration}

Before starting the range-of-motion exercise game, users are given in-game instructions to stretch their neck as far as they can in six different directions (up, down, left, right, top left, and bottom right) and confirm these positions using the `A' button of the right controller. Coordinates of the calibration points (see Fig.\ref{fig:teaser}A) are recorded corresponding to the forward vector of the virtual headset when users stretch their neck, which provide the personalized range-of-motion targets during the game.

\subsubsection{Performance Score}
\label{sec:performance_score}
A performance score feature was implemented to provide users with feedback on their movement accuracy during the exercise, which is also valuable for tracking the quality and progress of the rehabilitation by the physicians. In addition, this also forms the basis for baseline \textbf{score-based gamification} for this study. Specifically, we measure how close the user tracks the moving starlight target at an instance, by computing the angle $\theta$ between the forward head pointing vector and the pointing vector from the virtual headset to the target. Such angle has a range of 0$\sim$180 degrees, and is computed every  $0.1$\,s starting from \textit{Calibration Point 1}. For user-friendly interpretation, we convert this head tracking discrepancy into an instance-based score, with a different treatment (\textit{accumulation} vs. \textit{penalization}) for the ``moving phase" and ``fixation phase" due to their distinct requirements of actions. 

\noindent
Specifically, in the ``moving phase", where the neck moves smoothly, the instance score accumulates and is computed as:

\begin{equation}
\Delta S_{moving} = \left(\frac{180^\circ - \theta}{180^\circ}\right)^8/8
\end{equation}
At the ``fixation phase", as the duration can be extended indefinitely when the user looks away from the target ($\theta>10^\circ$), a penalizing score will be computed instead: 
 
\begin{equation}
\Delta S_{fixation} = -[1 - \left(\frac{180^\circ - \theta}{180^\circ}\right)^8]/8
\end{equation} 

\noindent
Here, we used an exponent of 8 (empirically determined) to help set a tolerance of relatively small errors while heavily penalizing higher errors (e.g., $\theta=30^\circ$). However, this factor can be further adjusted with richer physiological data based on the recommendations of clinicians and patient conditions. For each movement segment (one ``moving phase"+ one subsequent ``fixation phase"), we accumulate the instance scores over time as $S_{segment}$. To keep the cumulative score for an exercise positive, a minimum of \textit{four points} per segment is awarded: 
\begin{equation}
S_{segment} = \max\left(\sum \Delta S_{moving}+\sum \Delta S_{fixation},\; 4\right),
\end{equation}
\noindent
The full score for the complete exercise is the sum of all 11 segment-wise scores. We perform rounding to show the scores as integers only. As a reference, a ``perfect accuracy" would lead to a maximum score of 238, while the lowest possible final score is 44.

\subsection{Gamification strategies}
While the general exercise routine remains the same, we formulate three different game designs, including \textbf{score-based gamification}, \textbf{spatiotemporal progression}, and \textbf{musical interaction}. With each paradigm, the user's movement will trigger different audio and visual reward and feedback, with the gameplay mechanics remaining the same. Here, the \textit{score-based gamification} will be used as a baseline to study the experience-based game designs. For a fair comparison, all games adopted the same virtual environment of the tropical island (near the beach) along with the associated environmental sound, and the visual elements during gameplay. For both \textit{score-based gamification} and \textit{musical interaction}, we fix the user at the same location and view point, and the time of the environment is set at an eternal sunset to trigger a sense of inner peace. For all three strategies, the final performance scores were computed. While segment-wise scores are shown constantly for \textit{score-based gamification} as a key feature, they are not shown in the two experience-based game designs, where implicit reward/feedback driven by movements is given instead. The details of the three paradigms are described in the following sections.

\subsubsection{Paradigm 1: Score-based gamification}

As shown in Fig. 1B, in this baseline paradigm, the user experiences the game in a fixed position on a rock located in the island's beach at sunset. We incorporated three main elements for conventional score-based gamification design for the ROM exercise. First, the full performance score is displayed on top of the screen with continuous updates as the game progresses. Upon completing each exercise segment, the segment-wise score will be shown right beside the starlight, Second, along with the segment-wise score, on-screen messages (e.g., "Great job!") are also displayed to motivate the users for completing the exercise. Finally, to keep the user engaged in the score-based paradigm, we used the royalty-free track \textit{Funky Dude} \textcolor{black}{from a free Unity store asset} as background music, which is blended with the natural sounds of the environment. The inclusion of this background track is to replicate the typical upbeat audio experience commonly found in engaging exercise games.

\subsubsection{Paradigm 2: Spatiotemporal progression}

\noindent
\textbf{Design principles}: Different from score-based gamification, the ``spatiotemporal progression" paradigm (see Fig. 1B) couples the completion of each exercise segment for the user with the progressive transitions in the virtual environment, including time of day (time acceleration in one exercise iteration), weather, and the locations/view points in the scene. This design is grounded in the established principles of the Stress Recovery Theory (SRT) \cite{ulrich1991stress,bratman2015nature}, Attentional Gate Model \cite{zakay1995attentional}, and Embodied Cognition \cite{barsalou2008grounded}. Specifically, SRT and Attention Gate Model support the application of dynamic natural environments and experience of compressed time progression for stress reduction, respectively. The peaceful tropical beach and accelerated temporal warping echo these principles. In addition, according to Embodied Cognition, which states that perception and cognition arise through bodily interaction with the environment, our design to couple physical movement with evolving virtual environment not only offers an implicit reward for the exercise, but also fostering relaxation and engagement.

\noindent
\textbf{Gameplay}: At the start of the ROM exercise, the user is placed on a rock in the ocean facing the island. The initial time of day is nighttime and it transitions gradually to sunrise by the end of the first iteration. Weather conditions alternate between rainy, cloudy, and clear states throughout the game. The time and weather shifts are triggered gradually upon the completion of one segment. During the second exercise iteration, the time of day gradually transitions from sunrise to sunset. At the end of the first round, where the user returns to the neutral neck position (Calibration point 6, Fig. 1A), they are relocated from the ocean rock to a rock on the island's beach. This relocation is implemented as a smooth teleportation transition: the environment first gradually fades to a solid gray color briefly, then the user is repositioned through teleportation seamlessly while the environment is still faded, and finally the environmental view of the new location on the island fades in. This approach avoids motion sickness that typically occur in abrupt teleportation in VR, and no user experienced motion sickness due to this. Using the same method, the user is returned back to the initial rock on the ocean at the end of the second iteration of the exercise, allowing the user to enjoy the view from the ocean upon completing the exercise. The final score is made visible to the user after completing the two iterations of the exercise.

\subsubsection{Paradigm 3: Musical interaction}

\noindent
\textbf{Design principles}: With a fixed view on the beach at sunset, the musical interaction paradigm introduces musical rewards for the user by linking neck motions in the ``moving phase" with the triggering of musical notes on top of an ambient background music. Our game design follows the the principles of Auditory-Motor Integration \cite{grahn2007rhythm}, which shows that coupling body movement with musical sound engages sensorimotor networks for emtion regulation, and Music Emotion Theory \cite{juslin2011handbook}, where musical stimuli can trigger neurochemical changes that inhibit the amygdala and reduce stress hormones like cortisol. In rehabilitation contexts, this movement–music coupling provides immediate multi-sensory feedback that enhances engagement, reinforces motor actions, and supports relaxed, rhythmically guided participation during ROM exercises.

\noindent
\textbf{Generative AI for music composition}:
As an exploration, we introduce AI generated music tracks to the audiovisual environment. \textit{First}, we used the multi-modal  \textit{Google FX} tool to create the relaxing ambient background music. Specifically, we used the following prompt: ``\texttt{zen, deep focus, meditation, minimalistic, no high-pitch tones, no hissing sound, no instruments, alpha wave}". In total, \textit{Google FX} generates three audio tracks of 30 seconds each that are intended to follow one another. We modified the tracks so they would seamlessly loop. \textit{Second}, we used Gemini 2.5 to create second level of decorative musical note sequences to be triggered via motion and blended with the alpha-wave ambient music to form a full composition. For each exercise segment, eight notes spaced by 1.5 seconds at the expected pace of the starlight target will be played while rests (empty notes) are allowed in the music sequence. We supplied these specifications as prompts to Gemini to obtain the compositions for 11 exercise segments in the game. Then, we used the \textit{pydub} audio library and Gemini (for sound texture search) to produce the synthesized bell-like sound for the music notes, as well as adding postprocessing, including reverberation, bandpass filtering, echo, and fades. We adjust the volumes of the alpha-wave ambient music, second layer of bell-like notes, and the environmental soundscape to optimize the auditory experience.

\noindent
\textbf{Gameplay}: During the game, the user will follow the starlight target with their neck motion. Along the trajectory of each segment, the bell-like music notes will be played sequentially on top of the ambient music. For this paradigm, the visual feedback of the starlight target was also modified. Instead of emitting a large burst of particles when the fixation at a calibration point is completed, the target produces pulses in random colors which match the notes in the musical sequences. As pulses coincide with musical notes in the sequence, this gives the impression the target is "singing" in response to the musical notes being played.

\section{User Study Setup}

\subsection{Participant Recruitment}
This study was approved by the local ethics committee of our institute. Upon informed consent, we recruited 20 participants (age = 28.1±4.3 years, 9 females \& 11 males) who had never been clinically diagnosed with cervical disorders or persistent neck pain (pain lasting longer than 3 months). To further characterize the participant cohort, we administered a short questionnaire regarding 1) the Numerical Pain Rating Scale (NPRS) score (scale 0$\sim$10) for neck pain experienced during the last 24 hours, 2). any neck problems (e.g., muscle stiffness, soreness, or pain) during the previous month, 3) familiarity with VR technology and exercise-related digital applications, and 4) their frequency of physical activity. Among these participants, 6/20 reported neck problems within the past one month, and in the previous 24 hours before the study, 5/20 showed a mild degree of neck pain 
(NPRS = 1$\sim$3, mean=1.88). \textcolor{black}{Note that these two figures overlap, with 14/20 reported no neck pain. 5/6 of participants who had neck pain were female.} Regarding familiarity with VR, 14/20 reported as being at least somewhat familiar with VR. Across all participants, five had previously received exercise rehabilitation therapy, and most of them (12/20) maintain a habit of doing physical exercise at least 3-5 times per week, with the rest (8/20) doing exercise 1-2 times per week.

\subsection{Study Design}
For the study, all participants were first given a brief PowerPoint presentation introducing the general knowledge of chronic neck pain and neck exercise rehabilitation, basics of the developed games, and the goals of the study. Then, they were instructed to complete a questionnaire collecting demographic and background information. For the study, the user performed ROM exercises under the three game designs described in Section 3.5. To mitigate the ordering effect, the order of the three paradigms was randomized for each participant. All participants completed the study with a sitting position in a steady chair with a solid back support. 

Before a VR exercise intervention, after remaining at rest for $\sim$5 min, the participants were asked to rate their stress level with the Subjective Units of Distress Scale (SUDS) \cite{wolpe1990practice} in the range of 0$\sim$100 (no distress to extreme distress). Then, we used the EmotiBit wearable sensor module to measure their pre-intervention physiological data (i.e., heart rate and electrodermal activity) for one minute. Afterwards, the participants can proceed with un-interrupted gameplay without verbal communications till completion. In each game, the participants were instructed to follow in-game instructions to set their personalized extreme range-of-motion through the calibration phase by stretching their neck comfortably in different directions. When users finished the calibration and entered the exercise phase, physiological data from the EmotiBit device was recorded again until the end of the exercise. To contrast the pre-intervention measure, we kept the last minute of the gameplay for analysis. Upon game completion, the participants were asked to rate their stress level again with the SUDS. In addition, they were asked to fill out additional questionnaires regarding the usability and user-experience related questions. Afterwards, they were instructed to have a 10-minute ``washout" period to continue with their normal activities outside the room for experiments, and then return to the next intervention. 

Upon completing all three game paradigms, the users were asked to provide rankings in terms of their personal preferences, anxiety reduction level, and suitability for long-term use, as well as freeform comments to justify their answers to all the items in the questionnaires.  

\begin{figure*}[htbp]
    \centering
    \includegraphics[width=\textwidth]{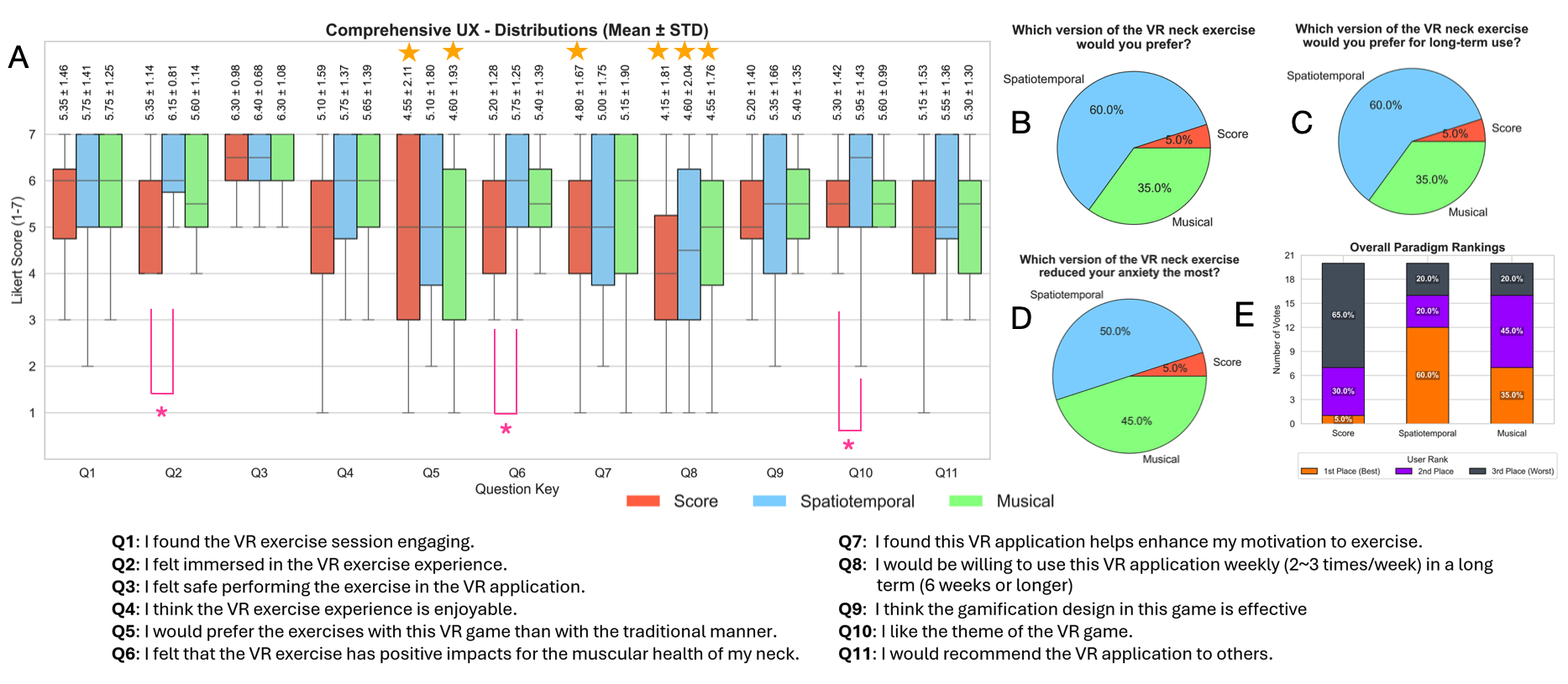}
    
    \caption{User experience (UX) question results. A. Boxplots of 11 UX questions with statistically significant pair-wise differences marked with asterisks (*) and detailed questions listed at the bottom. The mean$\pm$std values are annotated as well. The game paradigms' questions that received scores no different from a neutral score of 4 ($p>0.05$) are marked with yellow stars. B-D. pie charts for participant preferences: B. Participants' overall paradigm preference. C. Perceived paradigm that reduced anxiety the most. D. Paradigm preference for long-term use. E. Survey of participants' rankings of the three game paradigms.}
    \label{fig:graph-1}
\end{figure*}

\section{Evaluation Metrics}

From the user study, we collect semi-quantitative, quantitative, and qualitative metrics to test the three hypotheses set in Section~\ref{sec:hypotheses}.

\subsection{Semi-Quantitative and Freeform Questionnaire}
\label{sec:questionnaire}

Each of the three paradigms were evaluated by participants with semi-quantitative measures in the form of questionnaires. For each paradigm, we collected three types of semi-quantitative questions, including the System Usability Scale (SUS) \cite{brooke1996sus}, customized user experience questions (11 questions, see Fig. 3 for details) with a 7-point Likert scale (\textit{higher rating is preferred}), and the Subjective Units of Distress Scale (SUDS) \cite{wolpe1990practice} in the rating of 0$\sim$100 (no distress to extreme distress) before and after a VR intervention. The SUS is a commonly used instrument to assess software usability with ten questions, alternating between positive and negative questions, with answers ranging from 1 (strongly disagree) to 5 (strongly agree). From the questions, we calculate a total SUS score ranging from 0 to 100 by multiplying each question's score by 2.5.  A software system that receives a SUS score above 68 indicates good usability \cite{sauro2011measuring}. In addition, upon completing all three game paradigms, the participant was asked three additional questions to select their top choices for 1) their preferred game, 2) the game that reduced their anxiety the most, and 3) the preferred game for long-term use. 
Finally, participants were asked to rank the different paradigms, and we also invited them to give freeform comments regarding the positives and negatives of the games, as well as to offer suggestions for future improvements.

\subsection{Physiological Data}
\label{sec:physiological}

In addition to semi-quantitative questionnaires, we collected physiological data using the \textit{EmotiBit} wearable module \cite{montgomery2024validating} to further confirm the impact of different game designs for stress/anxiety reduction. Specifically, we used its photoplethysmography (PPG) sensor (sampling rate of 25 Hz) to obtain the PPG signals for deriving the heart rate data using EmotiBit's API software. In addition, we also used its Galvanic Skin Response (GSR) sensor to measure the Electrodermal Activity (EDA) at a sampling rate of 15 Hz. To gauge the sympathetic nervous system arousal rather than immediate reactions to specific short-term stimuli, we obtained the tonic component of the EDA signal (or called skin conductance level/SCL), by applying a low-pass filter (cutoff frequency = 0.05 Hz) on the raw EDA time serial. For both heart rate and tonic EDA metrics, we apply the 1.5 $\times$ inter-quartile range (IQR) rule to remove data outliers and then take the average over time. The mean pre-game measurements upon outlier removal are then compared with those taken during the last minute of the exercise by subtracting them to obtain a difference for each recording. As the values are not normally distributed, we then evaluate the medians of these differences and compare the resulting median values across paradigms. Note that for reduced anxiety/stress level, we expect a drop in the heart rate and tonic EDA measures.

\subsection{Exercise Performance Data}

As described in Section~\ref{sec:performance_score}, the performance score is derived from the angular alignment between the user's head pointing direction and the direction of the starlight target. During the gameplay, in addition to the scores, we also recorded the raw deviation angular error values within the VR software and output them in a .JSON file, along with the time stamps to reflect whether the angles were recorded during the ``moving phase" or ``fixation phase". In addition, we recorded the calibration points' positions relative to the user head's neutral position as these reflect the flexibility and neck muscle health of the user.

\subsection{Statistical Data Analysis}
For all continuous value metrics, we performed Shapiro-Wilk normality tests and found many don't follow normal distributions across the game paradigms. Therefore, when comparing the scores from the SUS and UX-specific questions across three game paradigms, Friedman tests were utilized, followed by post-hoc analysis (Bonferroni-corrected Wilcoxon signed-rank tests) for pairwise comparisons. For total SUS scores for each game paradigm, the Wilcoxon signed rank test was used to determine if the results were significantly different from 68. For each of the customized user experience (UX) questions, we compared the results to a neutral response (score = 4) to confirm the general attitudes of the participants, also with the Wilcoxon signed rank test. Here, A p-value $<$ 0.05 was used to indicate a statistically significant difference. For the freeform feedback comments, we performed a qualitative summary by identifying recurring themes in participants’ written responses (e.g., relaxing, engaging, easy to use). Mentions of these themes were counted across participants and grouped by the paradigm context to provide an overview of commonly reported perceptions.

\section{Results}
\subsection{Semi-quantitative Evaluations}

The participants assessed all three game paradigms with SUS scores and customized UX questions. For the SUS scores for user-friendliness, the \textit{Score-based gamification}, \textit{Spatiotemporal progression}, and \textit{Musical interaction} received 85.75 ± 12.46, 88.88 ± 11.57, and 87.75 ± 12.03, respectively. While no statistically significant difference was observed among them, all their SUS scores were well above the usability threshold of 68 ($p<0.05$).

On the other hand, the answers to the eleven UX questions are illustrated as boxplots in Fig. 3A. Among the questions, the answers to three questions (Question 2, 6, and 10) had statistically significant group-wise differences (Friedman tests, $p$-value $< 0.05$). Specifically, for \textbf{Question 2} (``I felt immersed in the VR exercise experience."), the \textit{Spatiotemporal progression} paradigm had the highest score (6.15 ± 0.81), followed by the \textit{Musical interaction} (5.60 ± 1.14), and lastly the conventional score-based one (5.35 ± 1.14). With post-hoc analysis, the pair of \textit{Score-based} and \textit{Spatiotemporal progression} paradigms show a significant difference ($p$-value = 0.02, \textcolor{black}{Cohen’s d = 0.605}). Similarly, for \textbf{Question 6} ("I felt that the VR exercise has positive impacts for the muscular health of my neck."), the \textit{Spatiotemporal progression} paradigm had the highest rating (5.75 ± 1.25), which is significantly higher than that of the \textit{Score-based gamification} ($p$-value = 0.01, \textcolor{black}{Cohen’s d = 0.666}). Finally, \textbf{Question 10 }("I like the theme of the VR game.") also followed the same respective trend, with the \textit{Spatiotemporal progression} version having the highest mean (5.95 ± 1.43), followed by the musical one (5.60 ± 0.99), and the conventional one (5.30 ± 1.42). Again, Spatiotemporal progression was rated significantly better than the score-based one ($p$-value = 0.04, \textcolor{black}{Cohen’s d = 0.347}). For every other questions that had no significant group-wise differences, the mean of the \textit{Score-based gamification} consistently scored lower than the others. Finally, when looking at individual questions, only \textbf{Question 5} (\textit{Score-based} and \textit{Musical Interaction}), \textbf{Question 7} (\textit{Score-based}), and \textbf{Question 8} (all paradigms) failed to obtain scores for some game paradigms above a neutral response (score = 4).

In Fig. 3B-3C, we show the participants' preference selection among the three paradigms, with similar trends across the three questions. Specifically, as a holistic impression, a majority of participants (60\%) preferred \textit{Spatiotemporal progression}, followed by \textit{Musical interaction} (35\%), and finally the traditional \textit{Score-based} version (5\%). The same results were obtained when participants were asked which version of the VR neck exercise they would prefer for long-term use. When asked which version reduced their anxiety the most, the experience-based designs are favored by 95\% of the participants. Furthermore, when participants were asked to rank all three paradigms, 65\% ranked the score-based paradigm as the worst. Though 35\% ranked \textit{Musical Interaction} as the best, 45\% (9/20) found this version to be the second best. Generally, experienced-based designs are preferred among the participants, with \textit{Spatiotemporal progression} having a slight edge over \textit{Musical interaction}.

Lastly, based on self-reported SUDS scores for the distress level, participants reported having decreased anxiety level after intervention across the three different paradigms. From the results (see Fig. 4C), \textit{Spatiotemporal progression} decreased the distress level by 0.80 ± 1.61. The \textit{Musical interaction} and \textit{Score-based} paradigms decreased distress levels by 0.70 ± 1.45 and 0.45 ± 1.85, respectively. However, no statistical significance was observed ($p$-value = 0.11) among them.

\begin{figure}
    \centering
    \includegraphics[width=0.95\linewidth]{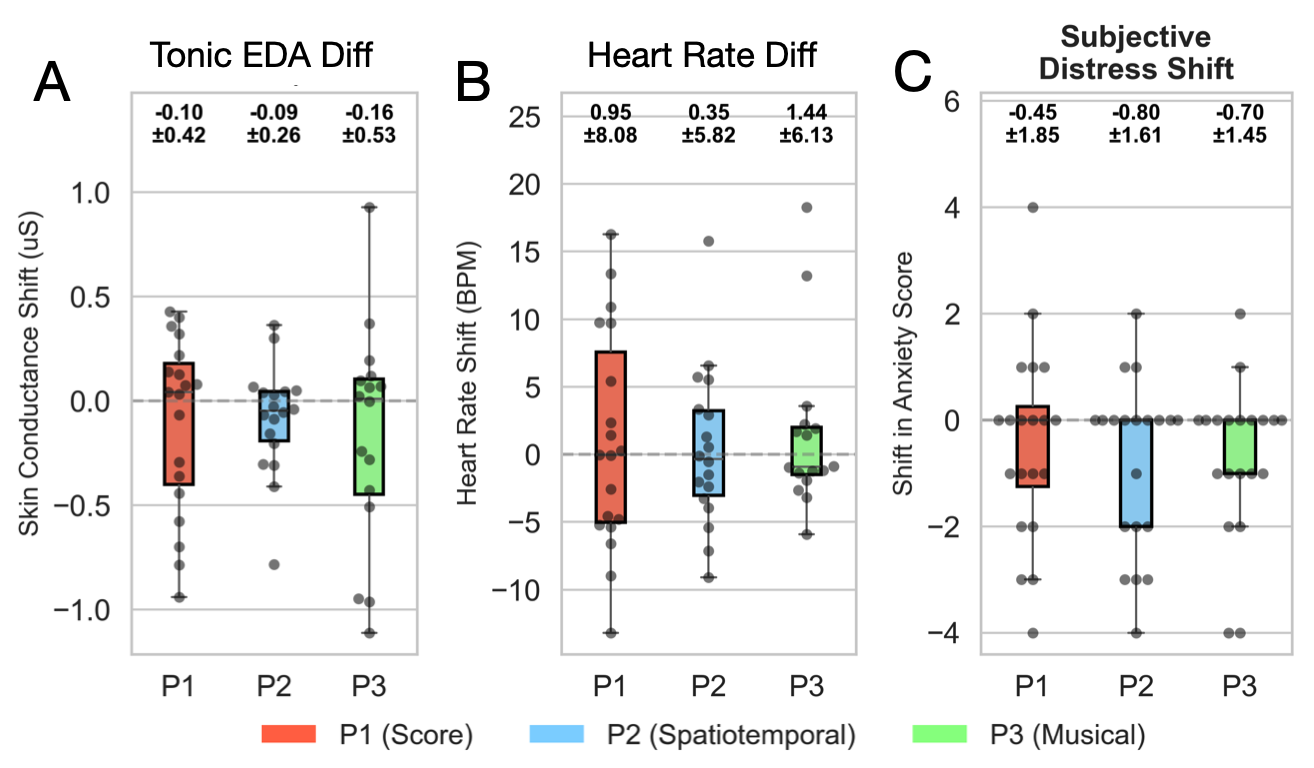}
    
    \caption{A and B. Heart rate and Tonic EDA shifts during the last minute of the gameplay from the before-intervention condition. C. Shifts in Subjective Units of Distress Scale from the VR games.}
    \label{fig:graph-2}
\end{figure}

\subsection{Quantitative Metrics}
For the recorded physiological data from the EmotiBit, using the 1.5 $\times$ IQR rule for outlier exclusion as stated in section~\ref{sec:physiological}, a Shapiro-Wilk test confirmed the data was normally distributed. The RM-ANOVA results showed no statistically significant difference in heart rate ($p$-value = 0.830) or Tonic Electrodermal activity ($p$-value = 0.132) between pre-game and during-game recordings. The box plots of the Tonic EDA shifts and heart rate shifts due to the game interventions are shown in Fig. 4A and 4B, respectively, In addition, the detailed values (mean$\pm$std), along with the statistical results are summarized in Table~\ref{tab:physiology}. On average, \textit{Musical interaction} offers the best anxiety-related physiological measurement reduction. Notably, \textit{Score-based gamification} raised the heart rate slightly on average, compared with experience-based designs.

\begin{table}[h]
\centering
\caption{Physiological Shifts (Mean $\pm$ STD)}
\label{tab:physiology}
\setlength{\tabcolsep}{3pt}
\begin{tabular}{@{}lccccc@{}}
\toprule
\textbf{Metric} & \textbf{Score} & \textbf{Spatiotemp.} & \textbf{Music} & \textbf{$F$} & \textbf{$p$} \\
\midrule
EDA ($\mu$S) & -0.10 $\pm$ 0.42 & -0.09 $\pm$ 0.26 & -0.34 $\pm$ 0.71 & 2.16 & .132 \\
HR (BPM) & 0.95 $\pm$ 8.08 & -0.56 $\pm$ 4.50 & -0.60 $\pm$ 2.53 & 0.19 & .830 \\
\bottomrule
\multicolumn{6}{@{}p{\linewidth}@{}}{\footnotesize \textit{Note.} Shift in physiological measurements between pre-exercise and during exercise. RM-ANOVA utilized.}

\end{tabular}
\end{table}

\subsection{Kinematic Accuracy}

\begin{figure}
    \centering
    \includegraphics[width=1\linewidth]{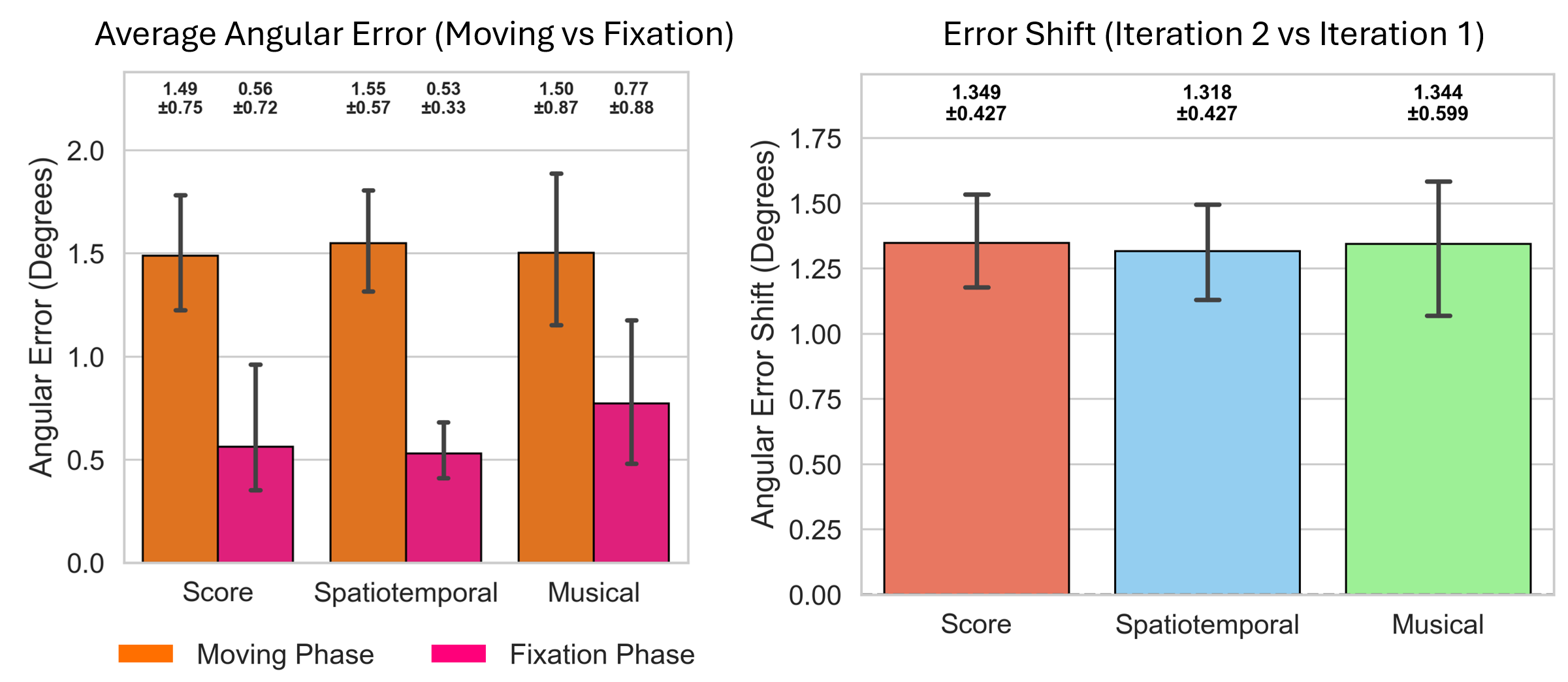}
    
    \caption{Kinematic accuracy with mean$\pm$std annotated. Left: average angular error in the moving and fixation phases; Right: Mean angular error shift from Iteration 1 to Iteration 2 (``Iter 2 - Iter 1").}
    \label{fig:graph-3}
\end{figure}

As an intuitive report for the performance quality, the final performance scores for all three paradigms are summarized in Table~\ref{tab:final_score}, and no statistical significant differences were observed. Overall, the participants performed well, with the reference of 238 as the maximum score.

\begin{table}[h]
\centering
\caption{Final Game Scores (Median [IQR])}
\label{tab:final_score}
\setlength{\tabcolsep}{3pt}
\begin{tabular}{@{}lccccc@{}}
\toprule
\textbf{Metric} & \textbf{Score} & \textbf{Spatio.} & \textbf{Music} & \textbf{$\chi^2$} & \textbf{$p$} \\
\midrule
Final Score & 220.47 [7.36] & 216.62 [10.72] & 219.34 [10.36] & 2.80 & .247 \\
\bottomrule
\multicolumn{6}{@{}p{\linewidth}@{}}{\footnotesize \textit{Note.} Data violated normality assumptions; therefore, a non-parametric Friedman test was utilized.}
\end{tabular}
\end{table}

When comparing the detailed angular-wise errors during target movements (``moving phase") compared to when the target is stationary (``fixation phase"), the average error is statistically higher when tracking a moving target ($p<0.05$) as shown in Fig.~\ref{fig:graph-3}. Interestingly, \textit{Musical interaction} demonstrated a statistically higher angular error from participants compared to \textit{Score-based gamification} during the ``fixation phase", as summarized in Table~\ref{tab:kinematic_errors}. As we expect good gamification strategies could encourage improved exercise performance over time, we computed the shifts of the mean angular error from the first to the second iteration of the exercises across three paradigms, with the results shown in Fig.~\ref{fig:graph-3}. Overall, all game paradigms resulted in a slight increase of angular errors in the second exercise iteration while the increases are on par with each other across the paradigms ($p>0.05$).

\begin{table}[h]
\centering
\caption{Kinematic Angular Errors (Median [IQR])}
\label{tab:kinematic_errors}
\setlength{\tabcolsep}{3pt}
\begin{tabular}{@{}lccccc@{}}
\toprule
\textbf{Metric} & \textbf{Score} & \textbf{Spatio.} & \textbf{Music} & \textbf{$\chi^2$} & \textbf{$p$} \\
\midrule
Moving Error ($^\circ$) & 1.27 [0.74] & 1.56 [1.02] & 1.31 [0.78] & 5.20 & .074 \\
Fixation Error ($^\circ$) & 0.37 [0.18] & 0.37 [0.30] & 0.50 [0.50] & 10.30 & .006* \\
\bottomrule
\multicolumn{6}{@{}p{\linewidth}@{}}{\footnotesize \textit{Note.} Metrics represent angular deviation from the target. Significant results ($p < .05$) are denoted by *. Non-parametric Friedman tests were utilized.}
\end{tabular}
\end{table}

\subsection{Freeform Questionnaire Results}
Fifteen participants provided written feedback describing positive and negative aspects of the system and suggesting potential improvements. \textcolor{black}{Note that the freeform feedback is optional for the participants}. Musical interaction was most consistently associated with relaxation, with 4/15 participants describing the musical interaction paradigm as calming or relaxing. Some noted that the music helped them relax, close their eyes, or even feel sleepy while performing the movements \textcolor{black}{(e.g., “Sound interaction made me feel sleepy and relaxed”). This suggests that the AI-generated music appeared effective in promoting relaxation and immersion, but it may have triggered drowsiness for some participants. As this could negatively affect the movement precision during the exercise. Future work could therefore investigate alternative music composition strategies that balance relaxation with maintaining alertness during rehabilitation exercises.} 

Spatiotemporal progression was frequently associated with engagement and curiosity about the evolving environment, with 5/15 participants commenting positively on the changing or evolving scene. However, 2/15 participants noted that the evolving environment could become overly relaxing or potentially lose novelty over time. \textcolor{black}{Example comments include ``It was interesting to see the world evolve as I completed tasks and it helped me stay engaged. Long term, I'm not sure I would prefer it over the sound interactions, because the novelty of the evolving world might wear off at some point." and ``It was grabbing my interest more when the ambient is changing, I wondered what will be the next change when I'm performing the movements". The feedback suggests that anticipation of future environmental changes contributed to engagement and relaxation by encouraging participants to remain curious about the evolving virtual world. If the novelty of dynamic environment falls behind the expectation, users may prefer the musical interaction approach instead. Future work should investigate strategies for maintaining novelty in long-term use, such as introducing a greater variety of environmental transitions and audiovisual events. }

Feedback regarding score-based gamification was more mixed. While 2/15 participants reported that seeing their score helped them focus on the task, another 2/15 participants indicated that the scoring system reduced the relaxing nature of the experience. Additionally, 2/15 participants described the score-based condition as boring compared to the other paradigms, and 2/15 participants reported that the audio used in the score-based condition was unpleasant. \textcolor{black}{For example comments include ``[It] was fun to see my score increase which ultimately helped me focus on the objectives of the task. My only concern is that because I could see my score it felt more like a game which made me feel more active rather than relaxed" and ``[It] was difficult to understand what got you more points; game would get boring with time". The user feedback suggests that the score-based approach may improve motivation and focus, but it could make the task boring and elevates the stress due to self-reflection on previous errors and future performance, especially when detailed scoring rules are not explained.}

\section{Discussion}

\subsection{Investigation of Hypotheses}
From the user study, our empirical results \textsl{partially supported} \textbf{Hypotheses 1} and \textbf{3} (\textbf{H1} \& \textbf{H3}), and fully support \textbf{Hypothesis 2} (\textbf{H2}). 

\subsubsection{H1. Anxiety Reduction}
According to user rankings (Fig.~\ref{fig:graph-1}D, participants perceived strongly that the two experience-based systems are superior in anxiety reduction. Physiological results obtained, shown in Table~\ref{sec:physiological}, indicate the largest decrease in Tonic EDA and heart rate with \textit{Musical interaction}. However, no statistically significant differences were found with the shifts in self-reported SUDS score or the physiological data from the VR intervention. As such, \textbf{H1} is partially supported, but still requires further study. Though the features in the experience-based game designs may help reduce the tonic EDA and heart rate, the physical exercise may act on the sympathetic nervous system to counteract on the physiological signals. In addition, for fair assessment, we presented the same immersive natural environment across three paradigms. As natural scenes have been shown to benefit mental well-being \cite{riches2023virtual}, it may have partially boosted the baseline \textit{Score-based} design in anxiety reduction. Compared between \textit{Spatiotemporal progression} and \textit{Musical interaction}, evidence demonstrates a slight preference towards the first. On the other hand, freeform feedback from participants suggested that the constant display of scores in \textit{Score-based gamification} may have been counterproductive for anxiety reduction, with negative terms for the paradigm more frequently reported. This interpretation is also consistent with the qualitative feedback, where several participants explicitly noted that the constant display of scores reduced the relaxing nature of the experience, while musical interaction was most frequently described as calming. From these findings, the use of experience-based designs may improve relaxation compared to the use of a conventional score display.

\subsubsection{H2. Performance quality}

From both the final performance scores and the recorded angle-wise errors, no significant differences were observed across the three paradigms. Despite not leveraging explicit rewards as in conventional \textit{Score-based gamification}, experience-based designs (\textit{Spatiotemporal progression} and \textit{Musical interaction}) can achieve similar exercise performance while offering potential additional benefit of anxiety reduction. However, based on the performance metrics, there are also some interesting subtle findings. First, as shown in Table 3, there is a slightly larger median angular error with \textit{Spatiotemporal progression} during the ``moving phase" than the other two (p = 0.074). We believe that the full virtual scene transition right after the ``fixation phase" could distract the user from the head tracking task. Second, in Musical interaction, we observed a higher angular error than the others at the ``fixation phase". Some participants mentioned that the AI-generated ambient music made them feel sleepy, and this could be more evident when holding the pose still. Overall, final scores showed no significant differences between paradigms. However, results suggest that spatiotemporal and alpha-wave features may influence performance, potentially due to increased relaxation and curiosity in experience-based paradigms, particularly for \textit{Spatiotemporal progression}. While the overall performance on average is not affected, features should be implemented while considering intended therapeutic outcomes. If timed correctly, experience-based features may not affect the performance and could \textcolor{black}{potentially} replace a score-based design.

\subsubsection{H3. User Experience}

Though there was a lack of statistically significant difference in most responses from the UX questions, the \textit{Score-based} paradigm consistently ranked worst in perceived user experience on average. Both experience-based paradigms ranked better in user experience and usability than \textit{Score-based gamification}, and in particular, the participants largely prefer them for long-term use, which is common for physical rehabilitation. As such, results observed tend to support \textbf{Hypothesis 3}.  From these results, we can observe that participants overwhelmingly preferred implicit in-game rewards/feedback that they could experience impressively instead of competitive performance scores, for the effect of reducing their anxiety more than the score-based version. This trend is further reflected in the qualitative feedback, where participants frequently described the spatiotemporal paradigm as engaging and the musical paradigm as relaxing, while the score-based condition received more mixed and negative comments. This suggests adherence to rehabilitation programs could \textcolor{black}{potentially} improve when using either experience-based designs. The spatiotemporal features may have the greatest effect, but both designs can be implemented conjunctly.

\subsection{Limitations and Future Works}

There are a few limitations in our current study. \textit{First}, as a preliminary investigation of the design strategies, we didn't include patient population in our participants as we would like to focus on the feasibility and acceptability of the experience-based design first. \textcolor{black}{As a preliminary study, using non-patients reduces pain-related complexities, including clinical variables (e.g., medication usage, pain severity) and the ability to sustain performance in three consecutive exercises. With a rising trend of the condition shifting to younger population \cite{murray2022prevalence}, our non-patient cohort (mean age = 28 yo) enables a more controlled assessment. However, one should note the insights gained from the current study are limited to the non-patient population.} In the next step, we will focus on patients with chronic neck pain by working closely with the clinicians to ensure safety and timely feedback. \textcolor{black}{This will help confirm the insights from this study for the target population.} \textit{Second}, we didn't incorporate any additional intervention, such as a Stroop test to raise the baseline stress level as our participants are primarily engineering graduate students, who generally experience high level of stress. However, in our cohorts, 40\% (8/20) of participants reported no distress or anxiety before the user study. As a result, the improvement of anxiety is less evident in these cases. \textit{Third}, although EDA and heart rate measures have been adopted to study the impact of emotional regulation, their measures can be affected by many internal and external factors and may contradict with subjective emotional perception \cite{barrett2007experience}. Future work could further integrate additional sensors, such as EEG to better characterize the impact of the experience-based design on anxiety reduction. \textcolor{black}{\textit{Fourth}, as the “Spatiotemporal progression” and “Musical interaction” designs focus on distinct sensory experiences (motion-environment/time interaction vs. motion-auditory interaction), we assess them separately to better understand their individual impacts in the current study. Based on the overall positive feedback for the two paradigms, we hypothesize that a new game design that combines them both could lead to additive benefits of anxiety reduction. However, some drawbacks may also occur in the context of exercise rehabilitation. For example, enhanced calming effects may lead to drowsiness that reduces the motivation of exercises for some users. On the other hand, for other users, the accumulative multi-sensory engagement may also increase the cognitive load, making it difficult to focus on the exercise instruction and goals. Both of these can reduce the precision of the movements and thus therapeutic benefits. We will study the potential joint effects of these two designs in our future investigation. \textit{Finally}, our study presents a new exploratory investigation of experience-based gamification design critical for rehabilitation exercise, so we didn't conduct family-wise correction for individual UX survey items. While the few statistical significant comparisons limit robust conclusions, this aligns with the exploratory nature of our study, which aims to identify emerging trends and establish potential directions to address the psychological aspect of VR-based rehabilitation exercises. We will further confirm the preliminary results in the future study with a larger patient cohort and additional physiological measures (e.g., EEG).}

\section{Conclusion}
In this \textcolor{black}{exploratory study}, we investigated novel experience-based designs for ROM exercises, common in the physical rehabilitation of chronic neck pain patients to address both the physical and psychological aspects of the therapy. Specifically, we designed the \textit{Spatiotemporal progression} and \textit{Musical interaction} strategies based on established principles in psychology and neuroscience for anxiety/stress reduction. By comparing against the conventional \textit{Score-based gamification}, empirical evidence from our study suggests the potential benefits of experience-based VR exercise game designs for neck pain rehabilitation \textcolor{black}{among a non-patient cohort}. With this preliminary investigation, further exploration based on a larger \textcolor{black}{patient group with chronic neck pain} would greatly enhance the understanding of multi-sensory interaction design and their application in rehabilitation exercises that also demand effective emotion regulation.

\acknowledgments{%
  This work was supported by the New Frontiers in Research Fund - Exploration grant (\# NFRFE-2024-00508).%
}

\bibliographystyle{abbrv-doi-hyperref}

\bibliography{template}

@Article{Kazeminasab:2022:NPG,
  author = {Shirin Kazeminasab and Seyed Aria Nejadghaderi and Parisa Amiri and Hossein Pourfathi and Mohammad Araj-Khodaei and Mark J. M. Sullman and Ali-Asghar Kolahi and Saeid Safiri},
  title = {Neck pain: global epidemiology, trends and risk factors},
  journal = {BMC Musculoskeletal Disorders},
  year = {2022},
  volume = {23},
  number = {1},
  pages = {26},
  month = {January},
  doi = {https://doi.org/10.1186/s12891-022-05157-y}
}

@article{elbinoune2016chronic,
  title={Chronic neck pain and anxiety-depression: prevalence and associated risk factors},
  author={Elbinoune, Imane and Amine, Bouchra and Shyen, Siham and Gueddari, Sanae and Abouqal, Redouane and Hajjaj-Hassouni, Najia},
  journal={The Pan African Medical Journal},
  volume={24},
  pages={89},
  year={2016}
}

@article{Steilen2014,
   author = {Steilen, D. and Hauser, R. and Woldin, B. and Sawyer, S.},
   title = {Chronic neck pain: making the connection between capsular ligament laxity and cervical instability},
   journal = {Open Orthop J},
   volume = {8},
   pages = {326-45},
   ISSN = {1874-3250 (Print)
1874-3250},
   year = {2014},
   type = {Journal Article}
}

@article{Barreto2019,
   author = {Barreto, T. W. and Svec, J. H.},
   title = {Chronic Neck Pain: Nonpharmacologic Treatment},
   journal = {Am Fam Physician},
   volume = {100},
   number = {3},
   pages = {180-182},
   ISSN = {0002-838x},
   year = {2019},
   type = {Journal Article}
}

@article{Himler2023,
   author = {Himler, P. and Lee, G. T. and Rhon, D. I. and Young, J. L. and Cook, C. E. and Rentmeester, C.},
   title = {Understanding barriers to adherence to home exercise programs in patients with musculoskeletal neck pain},
   journal = {Musculoskelet Sci Pract},
   volume = {63},
   pages = {102722},
   ISSN = {2468-7812},
   year = {2023},
   type = {Journal Article}
}

@article{asiri2021kinesiophobia,
  title={Kinesiophobia and its correlations with pain, proprioception, and functional performance among individuals with chronic neck pain},
  author={Asiri, Faisal and Reddy, Ravi Shankar and Tedla, Jaya Shanker and ALMohiza, Mohammad A and Alshahrani, Mastour Saeed and Govindappa, Shashikumar Channmgere and Sangadala, Devika Rani},
  journal={PloS one},
  volume={16},
  number={7},
  pages={e0254262},
  year={2021},
  publisher={Public Library of Science San Francisco, CA USA}
}

@article{hao2024virtual,
  title={Virtual reality training versus conventional rehabilitation for chronic neck pain: A systematic review and meta-analysis},
  author={Hao, Jie and He, Zhengting and Chen, Ziyan and Remis, Andr{\'e}as},
  journal={PM\&R},
  volume={16},
  number={10},
  pages={1143--1153},
  year={2024},
  publisher={Wiley Online Library}
}

@article{ye2023use,
  title={The use of virtual reality in the rehabilitation of chronic nonspecific neck pain: a systematic review and meta-analysis},
  author={Ye, Gongkai and Koh, Ryan GL and Jaiswal, Kishore and Soomal, Harghun and Kumbhare, Dinesh},
  journal={The Clinical journal of pain},
  volume={39},
  number={9},
  pages={491--500},
  year={2023},
  publisher={LWW}
}

@article{Orr2023,
   author = {Orr, E. and Arbel, T. and Levy, M. and Sela, Y. and Weissberger, O. and Liran, O. and Lewis, J.},
   title = {Virtual reality in the management of patients with low back and neck pain: a retrospective analysis of 82 people treated solely in the metaverse},
   journal = {Arch Physiother},
   volume = {13},
   number = {1},
   pages = {11},
   ISSN = {2057-0082 (Electronic)
2057-0082 (Linking)},
   url = {https://www.ncbi.nlm.nih.gov/pubmed/37194037},
   year = {2023},
   type = {Journal Article}
}

@article{Guo2024,
   author = {Guo, Q. and Zhang, L. and Han, L. L. and Gui, C. and Chen, G. and Ling, C. and Wang, W. and Gao, Q.},
   title = {Effects of Virtual Reality Therapy Combined With Conventional Rehabilitation on Pain, Kinematic Function, and Disability in Patients With Chronic Neck Pain: Randomized Controlled Trial},
   journal = {JMIR Serious Games},
   volume = {12},
   pages = {e42829},
   ISSN = {2291-9279 (Print)},
   year = {2024},
   type = {Journal Article}
}

@article{Global2024,
   title = {Global, regional, and national burden of neck pain, 1990-2020, and projections to 2050: a systematic analysis of the Global Burden of Disease Study 2021},
   journal = {Lancet Rheumatol},
   volume = {6},
   number = {3},
   pages = {e142-e155},
   ISSN = {2665-9913},
   year = {2024},
   type = {Journal Article},
   author = {GBD 2021 Neck Pain Collaborators}
}

@article{Chiu2024,
author = {Chiu, Maria and Tochilnikova, Elina and Harteveld, Casper},
title = {From Novelty to Clinical Practice: Exploring VR Exergames with Physical Therapists},
year = {2024},
issue_date = {October 2024},
publisher = {Association for Computing Machinery},
address = {New York, NY, USA},
volume = {8},
number = {CHI PLAY},
url = {https://doi.org/10.1145/3677068},
journal = {Proc. ACM Hum.-Comput. Interact.},
month = oct,
articleno = {303},
numpages = {29}
}

@article{yang2021understanding,
  title={Understanding the dark side of gamification health management: A stress perspective},
  author={Yang, Hualong and Li, Dan},
  journal={Information Processing \& Management},
  volume={58},
  number={5},
  pages={102649},
  year={2021},
  publisher={Elsevier}
}

@article{hoffmann2017gamification,
  title={Gamification in stress management apps: a critical app review},
  author={Hoffmann, Alexandra and Christmann, Corinna A and Bleser, Gabriele},
  journal={JMIR serious games},
  volume={5},
  number={2},
  pages={e7216},
  year={2017},
  publisher={JMIR Publications Inc., Toronto, Canada}
}

@inproceedings{hosseini2025evaluating,
  title={Evaluating the impact of immersiveness in virtual reality simulations on anxiety reduction for mri procedures: A preliminary study},
  author={Hosseini-Toudeshky, Hamideh and Seidnitzer, Sarah and Bickelhaupt, Sebastian and Harmouche, Rola and Kersten-Oertel, Marta},
  booktitle={2025 IEEE Conference Virtual Reality and 3D User Interfaces (VR)},
  pages={613--622},
  year={2025},
  organization={IEEE}
}

@article{barrett2007experience,
  title={The experience of emotion},
  author={Barrett, Lisa Feldman and Mesquita, Batja and Ochsner, Kevin N and Gross, James J},
  journal={Annu. Rev. Psychol.},
  volume={58},
  number={1},
  pages={373--403},
  year={2007},
  publisher={Annual Reviews}
}

@inproceedings{kim2025designing,
  title={Designing VR Music Game for Stress Reduction},
  author={Kim, Kirak and Kim, Hyojin and Choi, Youjin and Nam, Juhan and Lee, Jeongmi},
  booktitle={2025 IEEE Conference Virtual Reality and 3D User Interfaces (VR)},
  pages={678--685},
  year={2025},
  organization={IEEE}
}

@inproceedings{lecamwasam2023investigating,
  title={Investigating the physiological and psychological effect of an interactive musical interface for stress and anxiety reduction},
  author={Lecamwasam, Kimaya and Gutierrez Arango, Samantha and Singh, Nikhil and Elhaouij, Neska and Addae, Max and Picard, Rosalind},
  booktitle={Extended Abstracts of the 2023 CHI Conference on Human Factors in Computing Systems},
  pages={1--9},
  year={2023}
}

@article{zhang2026vr,
  title={VR Calm Plus: Coupling a Squeezable Tangible Interaction with Immersive VR for Stress Regulation},
  author={Zhang, He and Li, Xinyang and Zhou, Xingyu and Fu, Xinyi},
  journal={arXiv preprint arXiv:2602.05093},
  year={2026}
}

@article{zhang2026asafeplace,
  title={ASafePlace: User-Led Personalization of VR Relaxation via an Art Therapy Activity},
  author={Zhang, Chuyang and Yu, Bin and Wang, Yuchao and Yuan, Mansi and Wang, Wanqi and Je, Seungwoo and An, Pengcheng},
  journal={arXiv preprint arXiv:2602.01579},
  year={2026}
}

@article{riches2023virtual,
  title={Virtual reality relaxation for people with mental health conditions: a systematic review},
  author={Riches, Simon and Jeyarajaguru, Priyanga and Taylor, Lawson and Fialho, Carolina and Little, Jordan and Ahmed, Lava and O’Brien, Aileen and van Driel, Catheleine and Veling, Wim and Valmaggia, Lucia},
  journal={Social psychiatry and psychiatric epidemiology},
  volume={58},
  number={7},
  pages={989--1007},
  year={2023},
  publisher={Springer}
}

@article{xu2024effectiveness,
  title={Effectiveness of virtual reality--based well-being interventions for stress reduction in young adults: systematic review},
  author={Xu, Joy and Khanotia, Areej and Juni, Shmuel and Ku, Josephine and Sami, Hana and Lin, Vallen and Walterson, Roberta and Payne, Evelyn and Jo, Helen and Rahimpoor-Marnani, Parmin},
  journal={JMIR mental health},
  volume={11},
  number={1},
  pages={e52186},
  year={2024},
  publisher={JMIR Publications Inc., Toronto, Canada}
}

@article{savoric2025systematic,
  title={Systematic review: The impact of virtual reality interventions on stress and anxiety in intensive care units},
  author={Savoric, Tjasa and Aziz, Safwan and Ling, Ryan Ruiyang and Antlej, Kaja and Arnab, Sylvester and Subramaniam, Ashwin},
  journal={Journal of critical care},
  volume={90},
  pages={155164},
  year={2025},
  publisher={Elsevier}
}

@article{yang2025analgesic,
  title={Analgesic effects and neural oscillatory mechanisms of music-synchronized virtual reality intervention},
  author={Yang, Qi-Hao and Du, Shu-Hao and Tang, Le and Zhang, Yong-Hui and Wang, Xue-Qiang},
  journal={Journal of NeuroEngineering and Rehabilitation},
  volume={22},
  number={1},
  pages={202},
  year={2025},
  publisher={Springer}
}

@article{skiers2025portable,
  title={Portable Silent Room: Exploring VR Design for Anxiety and Emotion Regulation for Neurodivergent Women and Non-Binary Individuals},
  author={Skier{\'s}, Kinga and Pai, Yun Suen and Nakagawa, Marina and Minamizawa, Kouta and Barbareschi, Giulia},
  journal={IEEE Transactions on Visualization and Computer Graphics},
  year={2025},
  publisher={IEEE}
}

@article{hsieh2023effect,
  title={The effect of water sound level in virtual reality: A study of restorative benefits in young adults through immersive natural environments},
  author={Hsieh, Chung-Heng and Yang, Ju-Yuan and Huang, Chun-Wei and Chin, Wei Chien Benny},
  journal={Journal of Environmental Psychology},
  volume={88},
  pages={102012},
  year={2023},
  publisher={Elsevier}
}

@article{koelsch2014brain,
  title={Brain correlates of music-evoked emotions},
  author={Koelsch, Stefan},
  journal={Nature reviews neuroscience},
  volume={15},
  number={3},
  pages={170--180},
  year={2014},
  publisher={Nature Publishing Group UK London}
}

@book{juslin2011handbook,
  title={Handbook of music and emotion: Theory, research, applications},
  author={Juslin, Patrik N and Sloboda, John},
  year={2011},
  publisher={Oxford University Press}
}

@article{grahn2007rhythm,
  title={Rhythm and beat perception in motor areas of the brain},
  author={Grahn, Jessica A and Brett, Matthew},
  journal={Journal of cognitive neuroscience},
  volume={19},
  number={5},
  pages={893--906},
  year={2007},
  publisher={MIT Press}
}

@article{ulrich1991stress,
  title={Stress recovery during exposure to natural and urban environments},
  author={Ulrich, Roger S and Simons, Robert F and Losito, Barbara D and Fiorito, Evelyn and Miles, Mark A and Zelson, Michael},
  journal={Journal of environmental psychology},
  volume={11},
  number={3},
  pages={201--230},
  year={1991},
  publisher={Elsevier}
}

@article{barsalou2008grounded,
  title={Grounded cognition},
  author={Barsalou, Lawrence W},
  journal={Annu. Rev. Psychol.},
  volume={59},
  number={1},
  pages={617--645},
  year={2008},
  publisher={Annual Reviews}
}

@article{bratman2015nature,
  title={Nature experience reduces rumination and subgenual prefrontal cortex activation},
  author={Bratman, Gregory N and Hamilton, J Paul and Hahn, Kevin S and Daily, Gretchen C and Gross, James J},
  journal={Proceedings of the national academy of sciences},
  volume={112},
  number={28},
  pages={8567--8572},
  year={2015},
  publisher={National Academy of Sciences}
}

@article{zakay1995attentional,
  title={An attentional-gate model of prospective time estimation},
  author={Zakay, Dan and Block, Richard A and others},
  journal={Time and the dynamic control of behavior},
  volume={5},
  pages={167--178},
  year={1995}
}

@book{wolpe1990practice,
  title={The practice of behavior therapy},
  author={Wolpe, Joseph},
  year={1990},
  publisher={Pergamon press}
}

@article{sauro2011measuring,
  title={Measuring usability with the system usability scale (SUS)},
  author={Sauro, Jeff},
  year={2011}
}

@article{weber2020commercially,
  title={How commercially available virtual reality--based interventions are delivered and reported in gait, posture, and balance rehabilitation: a systematic review},
  author={Weber, Heather and Barr, Christopher and Gough, Claire and Van den Berg, Maayken},
  journal={Physical therapy},
  volume={100},
  number={10},
  pages={1805--1815},
  year={2020},
  publisher={Oxford University Press}
}

@article{vlaeyen2016fear,
  title={The fear-avoidance model of pain},
  author={Vlaeyen, Johan WS and Crombez, Geert and Linton, Steven J},
  journal={Pain},
  volume={157},
  number={8},
  pages={1588--1589},
  year={2016},
  publisher={LWW}
}

@article{montgomery2023introducing,
  title={Introducing {EmotiBit}, an open-source multi-modal sensor for measuring research-grade physiological signals},
  author={Montgomery, Sean M. and Nair, Nitin and Chen, Phoebe and Dikker, Suzanne},
  journal={Science Talks},
  volume={6},
  pages={100181},
  year={2023},
  publisher={Elsevier},
  doi={10.1016/j.sctalk.2023.100181}
}

@article{montgomery2024validating,
  title={Validating {EmotiBit}, an open-source multi-modal sensor for capturing research-grade physiological signals from anywhere on the body},
  author={Montgomery, Sean M. and Nair, Nitin and Chen, Phoebe and Dikker, Suzanne},
  journal={Measurement: Sensors},
  volume={32},
  pages={101075},
  year={2024},
  publisher={Elsevier},
  doi={10.1016/j.measen.2024.101075}
}

@inproceedings{brooke1996sus,
  author    = {John Brooke},
  title     = {SUS: A "Quick and Dirty" Usability Scale},
  booktitle = {Usability Evaluation in Industry},
  editor    = {Patrick W. Jordan and Bruce Thomas and Bernard A. Weerdmeester and Ian L. McClelland},
  publisher = {Taylor and Francis},
  year      = {1996},
  pages     = {189--194}
}

@article{murray2022prevalence,
  title={The prevalence of chronic pain in young adults: a systematic review and meta-analysis},
  author={Murray, Caitlin B and de la Vega, Roc{\'\i}o and Murphy, Lexa K and Kashikar-Zuck, Susmita and Palermo, Tonya M},
  journal={Pain},
  volume={163},
  number={9},
  pages={e972--e984},
  year={2022},
  publisher={LWW}
}

@article{wang2026global,
  title={Global, Regional, and National Burden of Neck Pain Among Adolescents and Young Adults, 1990--2021, and Projection to 2035: A Trend and Cross-Country Inequality Study},
  author={Wang, Fei and Lu, Hao and Cao, Yu and Li, Dilu and Cao, Lingjie and Zhang, Hanqing and Chen, Zhiming},
  journal={Journal of Pain Research},
  pages={581177},
  year={2026},
  publisher={Taylor \& Francis}
}

@article{gao2023risk,
  author    = {Gao, Yifang and Chen, Zhiming and Chen, Shaoqing and Wang, Shizhong and Lin, Jianping},
  title     = {Risk factors for neck pain in college students: a systematic review and meta-analysis},
  journal   = {BMC Public Health},
  year      = {2023},
  volume    = {23},
  number    = {1},
  pages     = {1502},
  month     = aug,
  doi       = {10.1186/s12889-023-16212-7},
  publisher = {Springer Nature}
}

\appendix 
\crefalias{section}{appendix} 

\end{document}